\documentstyle[aaspp4,12pt]{article}
\def\clock{\count0=\time \divide\count0 by 60
     \count1=\count0 \multiply\count1 by -60 \advance\count1 by \time
     \number\count0:\ifnum\count1<10{0\number\count1}\else\number\count1\fi}

\begin{document}
\newcommand{\rg}{\sqrt{g}}
\newcommand{\etal}{{\it et al. }}
\newcommand{\Nabla}{\bigtriangledown}
\def\lesssim{\mathrel{\hbox{\rlap{\hbox{\lower4pt\hbox{$\sim$}}}\hbox{$<$}}}}
\def\gtrsim{\mathrel{\hbox{\rlap{\hbox{\lower4pt\hbox{$\sim$}}}\hbox{$>$}}}}
\let\la=\lesssim
\let\ga=\gtrsim
\newcommand{\dg}{\delta_g}
\newcommand{\dgr}{\delta^R_g}
\newcommand{\be}{\begin{equation}}
\newcommand{\ee}{\end{equation}}
\newcommand{\bz}{{\bf z}}
\newcommand{\bvl}{{\bf v}}
%\draft

%\begin{singlespace}
\title{Reconstructing Nonlinear Stochastic Bias from Velocity Space
Distortions}  
\author{Ue-Li Pen}
\affil{Harvard Society of Fellows and Harvard-Smithsonian Center
for Astrophysics}
\altaffiltext{1}{e-mail I: upen@cfa.harvard.edu}

\begin{abstract}

We propose a strategy to measure the dark matter power spectrum using
minimal assumptions about the galaxy distribution and the galaxy-dark
matter cross-correlations.  We argue that on large scales the central
limit theorem generically assures Gaussianity of each smoothed density
field, but not coherence.  Asymptotically, the only surviving
parameters on a given scale are galaxy variance $\sigma$, bias
$b=\Omega^{.6}/\beta$ and the galaxy-dark matter correlation
coefficient $r$.  These can all be determined by measuring the
quadrupole and octupole velocity distortions in the power spectrum.
Measuring them simultaneously may restore consistency between all
$\beta$ determinations independent of galaxy type.

The leading deviations from Gaussianity are conveniently parameterized
by an Edgeworth expansion.  In the mildly non-linear regime, two
additional parameters describe the full picture: the skewness
parameter $s$ and non-linear bias $b_2$.  They can both be determined
from the measured skewness combined with second order perturbation
theory or from an N-body simulation.  By measuring the redshift
distortion of the skewness, one can measure the density parameter
$\Omega$ with minimal assumptions about the galaxy formation process.
This formalism also provides a convenient parametrization to quantify
statistical galaxy formation properties.

\end{abstract}

\section{Introduction}

The measurement of the distribution of matter in the universe has been
one of the frontier goals of modern cosmology.  The correlation of
galaxies has been measured in several surveys and is known to
significant accuracy.  The abundance of data has led to many
theoretical challenges, especially for flat Cold Dark Matter (CDM)
cosmologies.   In the simplest models, galaxies are considered an
inbiased tracer of the mass.  Several different measurements of
velocities then allow us to measure the density of the total matter,
using either cluster mass-to-light ratios or pairwise velocities
(Peebles 1993).  Under these assumptions, one obtains values of the
density parameter $\Omega_0\sim 0.2$.  It is known, however, that
galaxies of different types correlate in different ways (Strauss and
Willick 1995).  Thus, not all galaxies can simultaneously trace the
mass, and it appears plausible that no galaxy type is a perfect tracer
of mass.  At this point, galaxy formation is not completely
understood.  Simulations of galaxy formation suggest that the galaxy
formation process is stochastic and non-linear (Cen and Ostriker
1992).  This complicates the derivation of $\Omega_0$ from dynamical
measurements.  Peculiar velocity fields in principle allow one to
measure the mass fluctuation spectrum (Strauss and Willick 1995), but
even here one must assume that statistical properties of galaxies,
such as the Tully-Fischer relation, do not depend on the local density
of matter.

Assumptions need to be applied to relate the distribution of visible
galaxies to that of the total matter.  The most popular of relations
has been to assume linear biasing, where the power spectrum of
galaxies $P_g(k)$ is related to the matter power spectrum $P(k)$ by
$P_g(k) = b^2 P(k)$, motivated by the peak biasing paradigm (Kaiser
1984).  The gravitational effects of dark matter lead to velocity
fields which amplify the redshift space power spectrum in the linear
regime.  By measuring the distortions and imposing a stronger
assumption $\delta_g=b\delta$, a parameter $\beta=\Omega^{.6}/b$ can
be constrained (Kaiser 1987).  Recent attempts to measure these have
resulted in a confusing picture, wherein the results depend strongly
on the galaxy types (Hamilton 1997), scale (Dekel 1997) and surveys.
A potential explanation for this state of confusion is that the linear
galaxy biasing model is too simplistic.  It has been proposed that
galaxy formation may be non-local (Heyl \etal 1995).  A general model
assuming only that galaxy formation is a stochastic process has been
proposed by Dekel and Lahav (1997).  Unfortunately, the most general
stochastic distribution introduces several new free functions, which
make an unambiguous measurement of the matter distribution seemingly
impossible.  In principle, higher order statistics (Fry 1994) can be
used to to break some of the degeneracies, but the application of the
full theory is still in its infancy.

Nevertheless, the fact that the correlation properties are distorted
in redshift space tells us that gravitational interactions are at
work.  The challenge is to translate these measured distortions into
conventional cosmological parameters.  An added complication arises
from the fact that the redshift distortions are more accurately
measured in the quasi-linear regime where non-linear corrections are
believed to be important (Hamilton 1997).  If we are given complete
freedom to create galaxies with any distribution we please, and to
allow these galaxies to have any form of correlation or
anti-correlation with the dark matter density, can we still say
something about the dark matter distribution?  Can we invoke more
observables to constrain the exhaustive class of galaxy distribution
models?  In this paper we present the framework for a formalism which
allows us to parameterize all galaxy distribution freedom into a
simple unified picture.  In section \ref{sec:pdf} we begin by
constructing a hierarchy of approximations.  The central limit theorem
argues that the density field of both the galaxy distribution and of
the matter distribution will tend to be Gaussian when smoothed on
sufficiently large scales.  Their correlation coeffient, however,
remains a free parameter.  For general random processes, a bivariate
Gaussian distribution will have an arbitrary correlation coefficient
$-1\le r \le 1$.  The distribution has three free parameters: two
variances and a cross-correlation coefficient.  We show in section
\ref{sec:z} how to determine all three using linear power spectrum
distortions.  The first deviation from Gaussianity is characterized by
skewness.  For a bivariate distribution, one has in general four
skewness parameters.  We show in section \ref{sec:pdf} how to reduce
this number systematically.  Once the power spectrum of the matter has
been determined, the skewness of the matter can be computed using
either second order perturbation theory (Hivon \etal 1995) or using
N-body simulations.  Skewness is also subject to redshift distortions,
which can be measured.  We show how these two additional observables
allows one to constrain $\Omega$ in section \ref{sec:z}.  Individual
galaxies may be a biased tracer of the local dark matter velocity
field (Carlberg \etal 1990).  But it is reasonable to expect that when
smoothed on sufficiently large scales, the averaged velocity field
will be unbiased.  A geometric interpretation of the joint
distribution function is given in section \ref{sec:coherent}.  To
assure accuracy, any parameters derived using this formalism must be
checked against an N-body simulation.  Such simulations also allow one
to probe further into the non-linear regime.  We describe in section
\ref{sec:mock} how to build mock galaxy catalogs which obey the formal
expansion presented in this paper.

\section{Galaxy Distribution}
\label{sec:pdf}
\newcommand{\bx}{{\bf x}}

We will analyze  the density field smoothed by a tophat
window $W^R(r)$ on some scale $R$, which is zero for $r>R$ and $3/4\pi
R^3$ elsewhere.  The perturbation variable $\delta \equiv
(\rho-\bar{\rho})/\bar{\rho}$ is convolved to obtain the smoothed
density field $\delta^R = \int d^3\bx' \delta(\bx') W(|\bx-\bx'|)$.  The
galaxy density field will be described with a subscript $_g$, and we
similarly derive a smoothed galaxy density field $\delta^R_g$.  Only
the galaxy field is directly observable.  We do not have an exact
theory about how and where the galaxies formed, nor how they relate to
the distribution of the dark matter.  The smoothed galaxy density field
$\delta^R_g(\bx)$ could be a function of not only $\delta^R(\bx)$, but
also of its derivative, and possibly long range influences.  Even
non-gravitational effects may play a role.  Quasars exhibit a
proximity effect, which may influence galaxy formation.  If we
consider all these variables hidden and unknown, we must prescribe the
density of the galaxies $\delta^R_g$ as an effectively stochastic
distribution relative to the dark matter $\delta^R$.

In the most general galaxy distribution, we need to specify the
joint probability distribution function (PDF) $P(\delta^R_g,
\delta^R)$. 
At lowest order, we will approximate these to be Gaussians with some
finite covariance
\begin{equation}
P(\dgr,\delta^R) = \frac{\sqrt{ac-b^2}}{\pi}\exp \left(-\frac{a
(\dgr)^2}{2} + b\dgr\delta^R - \frac{c (\delta^R)^2}{2}\right).
\label{eqn:gal1}
\end{equation}
The parameters $a,b,c$ are related to the traditional quantities by a
change of variables: the variance of the matter $\sigma^2
= \langle(\delta^R)^2\rangle = \int\int (\delta^R)^2 P d\delta^R d\dgr =
a/(ac-b^2)$ and that of galaxies $\sigma_g^2 = c/(ac-b^2)$.  The third
free parameter is the covariance, given by the correlation coefficient
$r=\langle\dgr \delta^R\rangle/\sigma_g\sigma = b/\sqrt{ac}$.  The traditional
{\it bias} is defined as $b_1 = \sigma_g/\sigma = \sqrt{c/a}$.  The
power spectrum of the galaxies is then related to the power spectrum
of the dark matter by $P_g(k)=b_1^2P(k)$.  To simplify the notation,
we will from now on implicitly assume that all fields are smoothed at
scale $R$ unless otherwise specified, and drop the superscript $^R$ on all
variables.

The analysis of this paper relies on Equation (\ref{eqn:gal1}) being a
good approximation to the distribution of galaxies and dark matter
when smoothed on large scales.  A sufficient condition for this to
hold is if the central limit theorem applies.  Since we are
considering the statistics of density fields averaged over some
region, the central limit theorem will generally apply if the
smoothing region is larger than the scale of non-locality of galaxy
formation or non-linear gravitational clustering.  There are some
notable exceptions, however.  1. The dark matter power distribution
may be non-Gaussian even when the fluctuations are linear.  This might
be true, for example, in topological defect theories of structure
formation (Gooding \etal\ 1992) where the smoothed density field on
any scale may be non-Gaussian.  Such theories present great challenges
to many attempts to measure cosmological parameters, including for
example cosmic microwave background measurements (Pen, Seljak and
Turok 1997).  2. The galaxy fluctuations could depend non-linearly on
very large scale effects, for example external gravitational shear
(van de Weygaert and Babul 1994).  This could avoid the central limit
theorem due to non-locality.

Even if mild non-Gaussianity is present, we will argue below that the
problem remains tractable.  Conversely, there is every reason to
believe that on small scales, the distributions of both the galaxy and
the dark matter fields are significantly non-Gaussian.  We will show
in section \ref{sec:z} that mild non-Gaussianity actually allows us to
break the degeneracy between the bias factor $b_1$ and $\Omega$.  We
now turn to the next moment of the distribution, the skewness
$\langle\delta^3\rangle/\sigma^3$.  In principle one needs to specify
four such 
independent moments $\langle\delta^{3-i}\dg^{i}\rangle$ for $0\le i\le
3$.  A coordinate transformation simplifies Equation (\ref{eqn:gal1})
if we define $\sqrt{2a} \delta_g \equiv (u+v)/\sqrt{1-r}$, $\sqrt{2c}
\delta\equiv(u-v)/\sqrt{1-r}$, and $w^2\equiv(1-r)/(1+r)$.  $u$ is the
variable with unit variance along the joint distribution of the
galaxies and dark matter, while $v$ has variance $w^2$ and measures
their mutual deviation.  In this rotated frame, $u$ and $v$ are
uncorrelated.  In order to model all relevant terms, we
apply a general Edgeworth expansion about the Gaussian
(\ref{eqn:gal1}).  We recall (Kim and Strauss 1997) that the
coefficients of the two first order terms $u,v$ are zero since the
mean is by definition 0.  The three second moments are absorbed into
the definitions of $u,v$.  In principle one needs four third order
Hermite polynomials to describe the joint distribution
self-consistently at the next order.  But we can reasonably assume
that the galaxies are positively correlated with the dark matter
distribution, so $w \ll 1$.  In this case, the third order terms can
be rank ordered in powers of $w$ as $u^3,\ u^2v,\ uv^2,\ v^3$.  As we
will see below, it is necessary to retain the first two terms to model
second order perturbation theory.  We then neglect the last two terms
because they depend on higher powers of the small parameter $w$, and
disappear completely in the limit that biasing is deterministic $r=1$.
The Edgeworth expansion then gives us the truncated skew distribution
\begin{equation}
P(u,v)=\frac{1+(u^3-3u)s+(u^2-1)v b_s/w^2}{2\pi w}
\exp\left(-\frac{u^2}{2}-\frac{v^2}{2w^2}\right). 
\label{eqn:gal2}
\end{equation}
The two new coefficients are the joint skewness parameter $s$, and the second
order bias $b_s$ which allows us to adjust the skewness of each
distribution independently.  The joint PDF will be discussed in more
detail in section \ref{sec:coherent} below.
The Taylor expansion for general biasing introduces a
quadratic bias at the same order as second order perturbation theory
(Fry 1994):
\begin{equation}
\delta_g =f(\delta) = b_1 \delta + b_2 (\delta^2-\sigma^2) + O(\delta^3).  
\label{eqn:bias}
\end{equation}
We will show in section \ref{sec:coherent} that in the stochastic
notation, $b_2 = 2b_sb_1/\sigma$.

We can now compute the basic relations needed for further calculations.  All
third order moments are uniquely defined
\begin{eqnarray}
\langle\delta^3\rangle &=& \sigma^3\left( \frac{1+r}{2}\right)^{3/2} (6s-6b_s)
\nonumber\\
\langle\delta^2\delta_g\rangle &=& \sigma^2\sigma_g\left(
\frac{1+r}{2}\right)^{3/2} (6s-2b_s) 
\nonumber\\
\langle\delta\delta_g^2\rangle &=& \sigma\sigma_g^2\left(
\frac{1+r}{2}\right)^{3/2} (6s+2b_s) 
\nonumber\\
\langle\delta_g^3\rangle &=& \sigma_g^3 \left( \frac{1+r}{2}\right)^{3/2}
(6s+6b_s)
\label{eqn:skew}
\end{eqnarray}

For an initially Gaussian matter distribution with a power-law power
spectrum $P(k)=k^n$, the {\it skewness factor} of the evolved matter
distribution 
is given by $S_3 \equiv \langle\delta^3\rangle/\sigma^4 = -(3+n)+34/7$
from second order perturbation theory
(Juskiewicz \etal 1993) which is exact for $\Omega=1$ and
depends only very weakly on $\Omega$ (Bouchet \etal 1992).  
We obtain one relation
\begin{equation}
\left(\frac{1+r}{2}\right)^{3/2}(6s-6b_s)=\sigma\left[\frac{34}{7}-(3+n)
\right]
\label{eqn:skewdm}
\end{equation}
for the 5 unknowns $\sigma,\sigma_g,r,s,b_s$.  We will use five
observational values to determine the remaining four galaxy
distribution parameters, as well as the cosmological parameter
$\Omega$.  These determinations will rely on the comparison between
redshift space and real space correlations.  The real space
correlation contains valuable information about the galaxy
distribution itself, while the redshift space distortions are a
consequence of the dynamics of the dark matter.

\section{Redshift space distortions}
\label{sec:z}
We first consider the measurement of redshift space distortions of the
variance or power spectrum in the presence of stochastic biasing.  At
first order, the two second order terms $s,b_s$ can be neglected.
The redshift space density of galaxies is affected by their velocities
through the Jacobian
$\rho_g(x) dx = \rho_g(x(z))(dx/dz) dz$ and linear perturbation theory
gives us
\begin{equation}
\delta_g(z) = \delta_g(x)+\delta(x)\Omega^{.6}\mu^2
\label{eqn:dgz}
\end{equation}
where $\mu=\cos(\theta)$ determines the 
angle between the wave vector $\hat{k_z}$ and the line-of-sight (Kaiser
1987) and we have made the distant observer approximation.
The power spectrum is the expectation value of the square of
the Fourier transform of (\ref{eqn:dgz}) and results in 
\begin{equation}
P_g(k_z,\mu) = P_g(k) (1+\beta^2\mu^4+2r\beta\mu^2)
\label{eqn:pgt}
\end{equation}
where $\beta=\Omega^{.6}/b_1$.
$P_g(k)$ is the undistorted power spectrum.
The Legendre relation
\be
P_l\equiv\frac{2l+1}{2}\int_{-1}^1 P_g(k_z,\mu) {\cal P}_l(\mu) d\mu
\label{eqn:leg}
\ee
where ${\cal P}_l(\mu)$ is the $l$-th Legendre polynomial (Hamilton
1997) allows us to obtain moments of the angular dependence of
(\ref{eqn:pgt}). One can in principle measure both $\beta$ and $r$ by 
measuring the quadrupole distortion $P_2$ and the next order
distortion $P_4$ separately.  
We can then solve for $r$ and $\beta$ using the following relations
\begin{eqnarray}
\frac{P_2}{P_0} =
\frac{\frac{4}{3}r\beta+\frac{4}{7}\beta^2}{1+\frac{2}{3}r\beta +
\frac{1}{5}\beta^2}
\nonumber \\
\frac{P_4}{P_0} =
\frac{\frac{8}{35}\beta^2}{1+\frac{2}{3}r\beta +
\frac{1}{5}\beta^2} \ \ .
\label{eqn:qpole}
\end{eqnarray}
With sufficiently large data sets, one can measure $r$ and $\beta$ as
a function of wavelength $k$.  An alternative approach would be to
measure $P_g$ from the angular correlation function, after which
determination of the monopole $P_0/P_g$ and quadrupole $P_2/P_g$ terms
would be sufficient.  Peacock (1997) compared $P_g$ derived from the
APM angular power spectrum to $P_0$ to determine $\beta_0=0.4\pm 0.12$
by setting $r=1$.  Allowing $r$ to vary will increase the inferred
value of $\beta$ for all such measurements.  In this case, the
relation between the actual value of $\beta$ for a given inferred
value of $\beta_0$ is $\beta=\sqrt{\beta_0^2+2\beta_0+r^2}-r$.  A
similar increase in $\beta$ for a given inferred $\beta_0$ using $r=1$
holds for quadrupole measurements using equation (\ref{eqn:qpole}).
If stochasticity has been neglected, all inferred values of $\beta$
are only lower bounds.

\newcommand{\bk}{{\bf k}}
The skewness can be obtained from the bispectrum $B_g(\bk_1,\bk_2,\bk_3)
\equiv \langle\dg(\bk_1)\dg(\bk_2)\dg(\bk_3)\rangle$.  Isotropy requires
$\bk_1+\bk_2+\bk_3=0$, using which we can compute the third moment
\begin{equation}
\langle\dg^3\rangle = \int \int \frac{d^3\bk_1 d^3\bk_2}{(2\pi)^3}
B_g(\bk_1,\bk_2,
-\bk_1-\bk_2) W^R(k_1) W^R(k_2) W^R(|\bk_1+\bk_2|).
\label{eqn:skewb}
\end{equation}
One can then measure the net skewness by inverting the angular
bispectrum to obtain the three dimensional bispectrum,
in analogy to the power spectrum from APM (Baugh and Efstathiou 1994). 
The skewness of the smoothed galaxy field determines the last equation in
(\ref{eqn:skew}).  The skewness of the dark matter field is determined
by the variance (\ref{eqn:skewdm}), allowing us to solve for both $s$
and $b_s$. 
Equations (\ref{eqn:skewdm},\ref{eqn:qpole},\ref{eqn:skewb}) 
allowed us to solve for all parameters of the stochastic
non-linear biased galaxy distribution model.  The final
goal is to break the degeneracy of $\beta=\Omega^{.6}/b$ to
determine $\Omega$ and $b$ independently.

In redshift surveys, the measured skewness is already distorted by
velocity distortions, but is nevertheless readily measurable (Kim and
Strauss 1997).  The second order perturbation theory calculation of
the skewness has recently been completed for both redshift space and
the real space (Juskiewicz \etal 1993, Hivon \etal 1995).  Their basic
result was that in the absence of biasing, the skewness factor $S_3$
(see above) is weakly dependent on $\Omega$, and is very similar in
real and redshift space in second order theory.  In these
calculations, $S_{3,z}$ and $S_3$ typically differ by a few percent
depending on $\Omega_0$.  A similar sized change occurs when the bias
parameter is changed.  Fortunately, the observations already allow
very accurate determinations, for example Kim and Strauss found
$S_3=2.93\pm 0.09$, where the errors are comparable to the expected
effect.  The actual redshift space skewness distortions quickly grow
significantly larger than the second order predictions.  For a
self-similar power spectrum with $n=-1$ and Gaussian filter
$\sigma=0.5$ Hivon \etal found $S_3=3.5$ and $S_{3z}=2.9$ using N-body
simulations, which is a very significant effect and many times larger
than current observational errors.  Second order perturbation theory
appears to systematically underestimate the redshift space skewness
distortions.  The real space skewness in simulations tends to be
higher than perturbation theory, while redshift space skewness tends
to be lower in simulations.  This trend suggests that direct N-body
simulations are necessary to quantify the effect of redshift space
distortions of skewness.  We will examine this strategy in section
\ref{sec:mock} below.

By measuring the skewness from solely angular correlation information
(Gazta\~naga and Bernadeau 1997) as well as from the redshift space
distribution, we obtain two measures of skewness which can be compared
to each other, from which one can solve for $\Omega$.  One could also
smooth the density field with an anisotropic window function, and
compute the dependence of the skewness of the smoothed density field
on the alignment of the window as was done for variance measurements
by Bromley (1994).  The formalism of Hivon \etal (1995) can then be
modified using these anisotropic window functions.  We will explore
this approach further in section \ref{sec:mock} below.  Unfortunately,
the second order perturbation theory redshift distorted skewness does
not have a simple closed form expression, and a numerical triple
integral must be evaluated for each specific set of choices of $n,\
\Omega,\ W(k,\mu)$.  Details of this procedure are given in Hivon
\etal (1995).

\section{Coherent Limit}
\label{sec:coherent}

A finite truncation of the Edgeworth expansion may result in a PDF
which is not positive everywhere.  For
small corrections $s\ll 1, b_s\ll w^2$, the PDF in Equation
(\ref{eqn:gal2}) remains positive in all regions where the 
amplitude is still large, and for practical purposes the negative
probabilities do not have a significant effect.  But it is possible to
leave the regime of small corrections.
When the corrections become large, the
PDF becomes negative when it still has a significant amplitude.  We
can, however, absorb the coefficient of 
$v$ into the exponent for small $b_s$
\be
P(u,v)=\frac{1+(u^3-3u)s}{\sqrt{2\pi}}
e^{-u^2/2}\times
\frac{\exp\left(-\frac{[v-(u^2-1)b_s]^2}{2w^2}\right)}{ w
\sqrt{2\pi}}.
\label{eqn:bs}
\ee
Equation (\ref{eqn:bs}) remains positive for all values of $b_s$.
The right term becomes a Dirac delta function in the 
limit $w\rightarrow 0$
\begin{equation}
P(u,v)=\frac{1+(u^3-3u)s}{\sqrt{2\pi}}
e^{-u^2/2}   \times \delta_D[v-(u^2-1)b_s].
\end{equation}
We then reproduce Equation (\ref{eqn:bias}) to leading order for the choices
$b_1=r \sigma/\sigma_g$ and $b_2 = 2b_sb_1/\sigma$.

\begin{figure}
\plotone{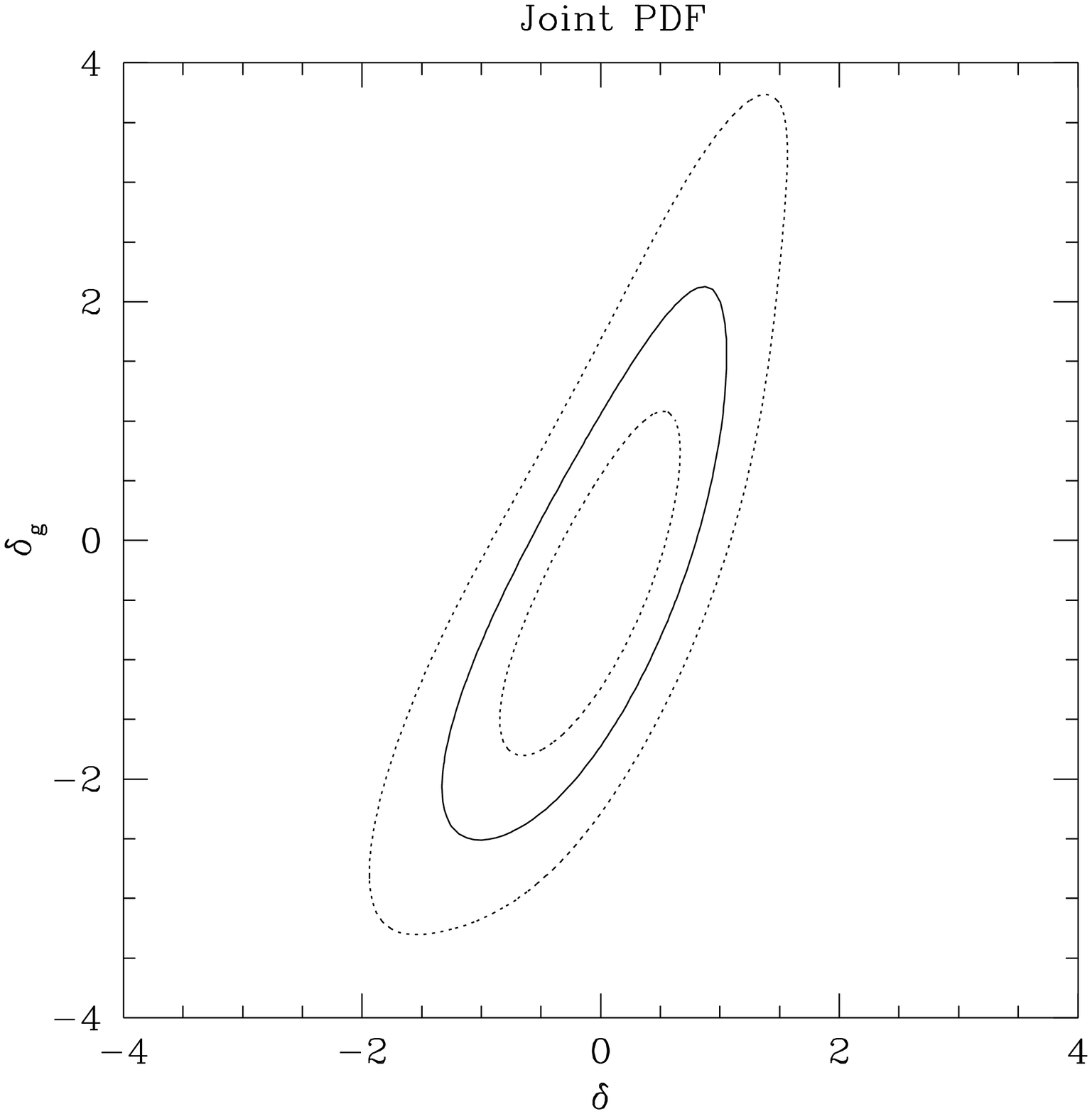}
\caption{Joint Probability Distribution Function Contours.  The
parameters for this plot are $\sigma=1$, bias $b=2$, correlation
coefficient 
$r=0.8$, skewness $s=0.05$ and non-linear bias $b_s=0.1$.  The solid
line is the contour at half central probability, while the dotted
lines are at 1/4 and 3/4.  The axes are in units of the dark matter
standard deviation.}
\label{fig:skew2d}
\end{figure}

\begin{figure}
\plotone{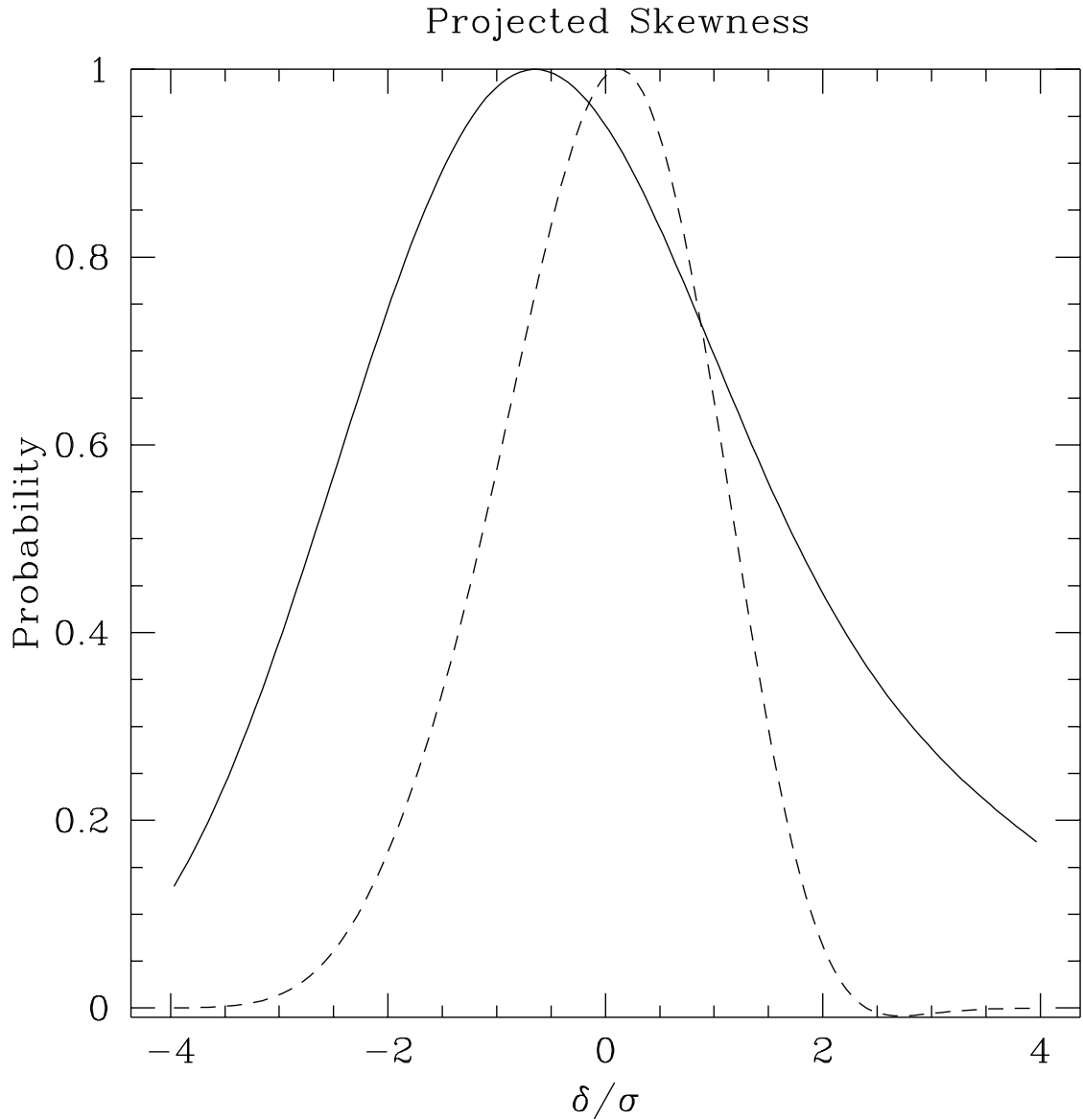}
\caption{The projection of Figure \protect{\ref{fig:skew2d}} onto the
galaxies (solid) and dark matter (dashed).  $s$ parametrizes their
common skewness, while each distribution's
skewness is proportional to $s\pm b_s$.  The units are in standard
deviations of 
the dark matter.  For our choice $b=2$ the galaxies have twice the
standard deviation of the dark matter.}
\label{fig:skew1d}
\end{figure}

It is instructive to understand the nature of the two skewness
parameters $s$ and $b_s$ graphically.  In Figure \ref{fig:skew2d} we
show the joint PDF.  The respective projections onto the galaxy
and dark matter PDF's is shown in Figure \ref{fig:skew1d}.
The projected PDF for the galaxies is
\be
P(\dg)=\frac{1+(\dg^3-3\dg)s_g}{\sigma_g\sqrt{2\pi}}
\exp\left(-\frac{\dg^2}{2\sigma_g}\right) 
\ee
where $s_g=[(1+r)/2]^{3/2}(s+b_s)$ and that for the dark matter is the
same with a change in sign of $b_s$.  For real surveys, one can
measure the residuals of the galaxy PDF after fitting a third order
Edgeworth expansion to determine the accuracy of the fit, with proper
account for noise (Kim and Strauss 1997).  The same can be done for the dark
matter by utilizing N-body simulations described in the next section. 

% place both figures here:

\section{Constructing Mock Catalogues}
\label{sec:mock}

Let us now examine how one can build galaxy catalogues from an N-body
simulation consistent with this Edgeworth expansion.  The purpose of
this exercise will be to test specific models against catalogs in the
non-linear regimes.  We will show a sample construction of a galaxy
density field smoothed by a window function which is consistent with
the stochastic non-linear biasing described above.  We can then
compite the likelihood function for the cosmological parameters for
any set of observations.  The ultimate test would be to recover the
same cosmological parameters and dark matter power spectrum using
galaxy types which are known to be biased relative to each other.

The first step is to calculate the bias function $b(k)$
in Fourier space by comparing the angular correlation function of the
survey to that of the simulation.
Statistical homogeneity and isotropy require that the bias is a
function of the magnitude of the wave-number only.
We will take the
density field of the simulation in Fourier space and produce the galaxy
field by scaling to the bias
\be
\dg(\bk)=b_1(|\bk|) \delta_{\rm DM}(\bk).
\ee
The mock density field is then convolved with the survey geometry and
projected onto an angular power spectrum $w(k)$.  We solve for the
bias function $b_1(k)$ by requiring the mock galaxy angular power
spectrum to agree with the observed angular power spectrum.  This
procedure has used no velocity information.  By repeating the
simulation many times with different random seeds, we can obtain the full
distribution of $b_1(k)$.

We have three remaining parameters $r,s,b_s$ which must be solved for
using velocity and skewness information.  Since the skewness of the
dark matter in the simulations is known, we have one constraint from
Equation (\ref{eqn:skew}), reducing them to two remaining degrees of
freedom.  We will use four observational quantities to constrain them:
two moments of the redshift space variance distortion, and the skewness
as well as its distortion.  While second order perturbation theory in
principle allows us to solve for $r,s,b_s$ and $\Omega$, its validity
quickly breaks down as one enters the non-linearly regime.  We must use
N-body simulations at this point to make quantitative comparison.  The
problem is now doubly overconstrained, allowing us to solve for two
free simulation parameters, for example $\Omega$ and the power spectrum
shape parameter $\Gamma$ by performing a sufficiently large number of
N-body simulations (Hatton and Cole 1997).  Velocity space distortions
can be measured using anisotropic smoothing windows (Bromley 1994)
$W^R(\mu)$.  The window function proposed by Bromley symmetrizes the
distribution and thus destroys skewness information.  Consider instead
an elliptical top-hat with a major-minor axis ratio of 2:1.

We pick a characteristic smoothing scale $R$ and smooth the
observation on that scale.  The trade-off occurs between picking large
$R$, which smoothes over large volumes and results in distributions
which are closer to Gaussian, or small $R$ which results in a smaller
cosmic variance and a stronger non-linear signal, but for which the
first three orders of our the Edgeworth expansion may be a poor
approximation for the true dark matter-galaxy joint distribution
function.  We first decohere the galaxy density field
$\rho_g$ by 
adding an independent random Gaussian galaxy field $\rho^N_g$ with
identical power spectrum weighted
by the correlation coefficient $r$:
\be
\rho'_g = r\rho_g+(1-r)\rho^N_g.
\ee
We have averaged the result of an N-body simulation with a random field
with identical (non-linear) power spectrum.
This maintains the shape of the power spectrum, but weakens the degree
of correlation between galaxy and dark matter fields.  It is no longer
true that $\langle\delta_g|\delta\rangle = b_1\delta$, but instead
$\langle\dg|\delta\rangle=b_1r\delta$. 
Second order bias is added by feeding the field through a quadratic
function
\be
\rho''_g = \rho'_g+\frac{2b_s}{\sigma_g}
\left[(\rho'_g-\bar{\rho_g})^2-b_1^2\sigma_g^2\right].
\label{eqn:mock}
\ee
Next we distort into velocity space as follows:  Each N-body particle
mass is multiplied by $\rho''_g/\rho_{\rm DM}$ and projected with its
velocity into a redshift coordinate system.  The window function is
applied in redshift space
\begin{eqnarray}
\rho_z(\bz)=\frac{\bar{\rho}}{\bar{n}}\sum_i m_i C[c\bz/H_0-(x_i+v_i^z)]
\nonumber \\
\rho^R(\bz)=\int \rho_z(\bz') W^R(|\bz-\bz'|,\mu) d^3\bz'.
\end{eqnarray}
$C(\bz)$ is the particle shape, which for Cloud-in-Cell mappings
(Hockney and Eastwood 1980) is the same shape as the grid cell.
$\bar{n}$ is the ratio of number of particles to the number of
gridcells and $m_i$ is the scaled particle mass.  $x_i$ is the
particle position, and $v^z_i$ is the line-of-sight component of the
particle velocity which affects the radial redshift position.

We now compare the statistics with the observed sample.  One computes
the variance $\sigma^2(\mu)=\int (\rho^R-\bar\rho^R)^2d^3\bz$ and
decomposes it into multipoles $\sigma^2(\mu)\sim\sigma_0+\sigma_2
P_2(\mu)+\sigma_4 P_4(\mu)$ as in Equation (\ref{eqn:leg}) and does
the same with the skewness $s_3(\mu)=\int (\rho^R-\bar\rho^R)^3d^3\bz$
where now $s_3\sim s_0+s_2 P_2(\mu)$.  These
$\sigma_2,\sigma_4,s_0,s_2$ are then compared with the values obtained
from the surveys.  A Monte-Carlo array of simulations provides the
full likelihood distribution of these variables, allowing us to test
consistency of each model with observations.  

This model of skew biasing allows us to discuss the systematic errors
in the measurement of pairwise velocity disperions (Guzzo \etal 1997).
Since pairwise galaxy velocities are measured in the non-linear
regime, the inferred mean galaxy velocities can not be directly
translated into mean dark matter pairwise velocity.  Decoherence, and
non-linear bias both introduce complex dependences in the conversion
from galaxy velocity to dark matter velocity.  Guzzo \etal (1997)
showed that the one dimensional pairwise velocity varies by galaxy
type from 345 to 865 km/sec.  We must keep in mind that each galaxy
type surely has different biasing and coherence properties.  The
pairwise velocities are typically measured at a separation of 1/h Mpc,
where the density field is strongly non-linear, and the Edgeworth
expansion may be a poor approximation to the actual joint galaxy-dark
matter distribution.  The mock catalog from N-body simulations
described above effectively provides a handle to probe the dynamical
properties of galaxies at larger separation, allowing us to separate
the distribution properties of galaxies from the dynamical aspects of
the dark matter.

\section{Conclusions}

The general stochastic galaxy biasing problem contains more free
parameters than can easily be measured in any galaxy redshift survey.
We have shown that using only linear perturbation theory we can
determine two parameters, the correlation coefficient $r$ and bias
parameter $\beta$ using the quadrupole and octupole distortions.  This
allows a reconstruction of the power spectrum $P(k)\Omega^{.6}$ as
well as determination of two galaxy formation parameters.  In the
plausible scenario that galaxies correlate strongly with the matter
distribution, only two free additional parameters $s,\ b_s$ need to be
introduced to quantify the skewness.  Second order perturbation theory
provides one linear constraint, and observations of the skewness of
galaxies determines the second.  By measuring the redshift distortions
of skewness we can in principle determine both the true underlying
dark matter power spectrum and the density parameter $\Omega$
independently.  This picture has incorporated both stochastic
correlation and second order non-linear bias.  We have shown how to
extend the {\it Ansatz} of the Edgeworth expansion to general problems
without relying on linear or second order theory.  In an N-body
simulation the same approach can be applied to directly compare
specific models to observations.  This also allows us to probe deeper
into non-linear scales.

{\sc Acknowledgements}.  I thank Ben Bromley, Uros Seljak and Michael
Strauss for helpful discussions.  This work was supported by the
Harvard Society of Fellows and the Harvard-Smithsonian Center for
Astrophysics.

\newpage

\end{document}